\documentstyle[12pt,epsf]{article}

\renewcommand{\bar}[1]{\overline{#1}}
\newcommand{\M}{{\cal M}}
\newcommand{\VEV}[1]{\left\langle{#1}\right\rangle}

\newcommand{\ie}{{\it i.e.}}

\newcommand{\half}{{1\over 2}}

\begin{document}
\begin{flushright}
SLAC--PUB--7839\\
June 1998
\end{flushright}
\bigskip\bigskip

\thispagestyle{empty}
\flushbottom

\begin{center}
{{\Large \bf Exact Light-Cone Wavefunction Representation\\[2ex]
of Matrix Elements of Electroweak Currents}\footnote{\baselineskip=14pt
     Work supported by the Department of Energy, contract
     DE--AC03--76SF00515.}}\\
\vspace{1.0cm}
Stanley J. Brodsky\\
{\it{Stanford Linear Accelerator Center\\
Stanford University,
Stanford, California 94309}}\\
{\rm and} \\
Dae Sung Hwang\\
{\it{Department of Physics, Sejong University, Seoul 143--747,
Korea}}
\end{center}

\vspace{2.0cm}
\begin{center}
Abstract
\end{center}
The matrix elements of electroweak currents which occur in exclusive decays of
heavy hadrons are evaluated in the nonperturbative light-cone Fock
representration. In general, each semileptonic exclusive decay  amplitude
receives two contributions, a diagonal $\Delta n = 0$ parton-number-conserving
amplitude and a
$\Delta n = -2$ contribution in which a quark and an antiquark from the initial
hadron Fock state annihilate to the leptonic current. The general formalism
can be used as a basis for systematic approximations to heavy hadron decay
amplitudes such as hard perturbative QCD contributions. We illustrate the
general formalism using a simple perturbative model of composite hadrons.  Our
analysis  demonstrates the occurence of ``zero-mode" endpoint contributions to
matrix elements of the ``bad" $j^-$  currents
in the Drell-Yan frame when $q^+ \to 0$.
\vfill
\begin{center}
(Submitted to Physics Letters B.)
\end{center}
\vfill


\newpage

\section{Introduction}

One of the most challenging problems
at the intersection of quantum chromodynamics and electroweak physics is the
evaluation of
exclusive decay amplitudes of heavy hadrons
such as the semileptonic decay
$B^0\rightarrow \pi^- \ell^+ \nu$.
The physics of such heavy hadron electroweak decays involve operator matrix
elements which depend in detail on the quark and gluon
composition of the initial state and final state hadrons.  Even the presence of
a heavy quark in the initial and/or final state does not simplify the complexity
of the QCD analysis,  since we must deal generally  with hadron wavefunctions
describing an arbitrary number of quark and gluon quanta.

In this paper we shall give formulas for the current matrix elements
$\VEV{A|J^\mu|B}$ describing general transition between hadrons $B$ and $A$.
The formulas are in principle exact, given the light-cone wavefunctions
of hadrons. Our results generalize the expressions for the elastic form factors
obtained by Drell and Yan \cite{DY,BD} and West \cite{West}. The underlying
formalism is the light-cone Hamiltonian Fock expansion in which  hadron
wavefunctions are decomposed on the free Fock basis of QCD. In this formalism,
the full Heisenberg current
$J^\mu$ can be equated to the current $j^\mu$ of the non-interacting theory
which in turn has simple matrix elements on the free Fock basis. In the case of
one-space and one-time theories,  such as collinear QCD \cite{AD},
the complete hadronic spectrum and the respective Fock state expansion
can be determined, at least  numerically,  using the DLCQ
(Discretized Light-Cone Quantization) method \cite {PB}.
Eventually full solutions can be envisaged for physical theories such as
QCD(3+1) using DLCQ,  Wilson's front-form formalism, lattice analyses,  and
other non-perturbative Hamiltonian methods.  For a review see Ref. \cite{BPP}.

An exact formalism provides the opportunity to make systematic
approximations and account for negelected terms.   For example, we can
identify the contributions to exclusive decay amplitude of heavy hadrons which
can be accounted for by hard perturbative QCD effects \cite{BHS}. On the other
hand, we also can identify specific physical mechanisms which are due to the
presence of higher Fock state non-valence configurations of the hadrons.

It is well known \cite{DY} that elastic form factors at
space-like momentum transfer
$q^2 = - Q^2 < 0$ are most simply  evaluated from matrix elements of the
``good"
current $j^+ = j^0 + j^z$ in the preferred Lorentz frame
where
$q^+ = q^0+ q^z = 0$.   The  $j^+$ current has the advantage that
it does not have large matrix elements to pair fluctuations, so that only
diagonal, parton-number-conserving transitions need to be considered.
The use of the  $j^+$ current and the $q^+ =
0$ frame brings out  striking advantage of the light-cone quantization
formalism: only diagonal, parton-number-conserving Fock state matrix elements
are required. However, in the case of the time-like form factors which occur in
semileptonic heavy hadron decays, we need to choose a frame with
$q^+ > 0$, where
$q^\mu$ is the four-momentum of the lepton pair.  Furthermore, in order to sort
out the contributions to the various weak decay form factors, we need to
evaluate the ``bad'' $-$ current $j^- = j^0 - j^z$ as well as the ``good''
current $j^+$.  In such cases
we will also require off-diagonal Fock state transitions; \ie\ the convolution
of Fock state wavefunctions differing by two quanta, a $q {\bar{q'}}$ pair.
The entire electroweak current matrix element is then in general given by the
sum of the diagonal $n \to n$ and off-diagonal
$n+1 \to n-1$ transitions. As we shall see, an important feature of a general
analysis is the emergence of
singular
$\delta(x)$ ``zero-mode" contributions from the off-diagonal matrix elements  if
the choice of frame dictates $q^+ = 0.$

\section{Matrix Elements of Electroweak Currents}

The light-cone Fock expansion is defined as the projection of an exact
eigensolution of the full light-cone quantized Hamiltonian on the
solutions of the free Hamiltonian with the same global quantum numbers.  The
coefficients of the Fock expansion are the complete set of $n$-particle
light-cone wavefunctions, $\{\psi_n(x_i, k_{\perp i}, \lambda_i)\}$.
The coordinates $x_i, k_{\perp i}$ are internal relative coordinates,
independent of the total momentum of the bound state, and satisfy $ 0 <
x_i < 1,\ \sum_i^n x_i = 1$ and $\sum_i^n  k_{\perp i} = 0_\perp$.
Here $x ={k^+\over  P^+} = {k^0 + k^3 \over P^0 + P^3}$ and we use the metric
convention
$a\cdot b={1\over 2}(a^+b^-+a^-b^+)-{\vec{a}}_{\perp}\cdot {\vec{b}}_{\perp}$.

\vspace{.5cm}
\begin{figure}[htb]
\begin{center}
\leavevmode
\epsfbox{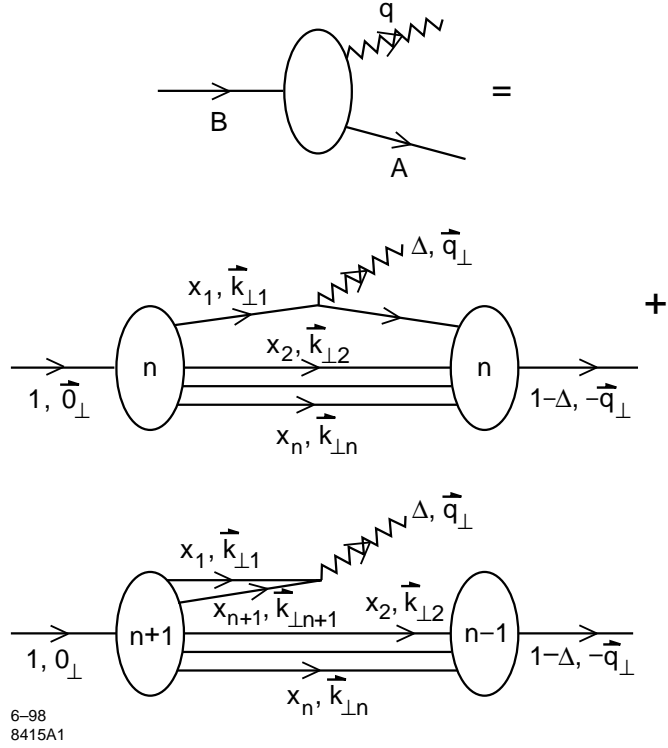}
\end{center}
\caption[*]{Exact representation of electroweak decays and time-like form factors in the
light-cone Fock representation.
}
\label{fig1}
\end{figure}

The evaluation of the semileptonic decay amplitude $B \to A \ell {\bar{\nu}}$
requires the matrix element  of the weak current between hadron states
$\VEV{B \vert j^\mu(0) \vert A}$.   (See Fig. 1.)
The interaction current then has simple matrix
elements of the free Fock amplitudes, with the provisal that all $x_i > 0.$  We
shall adopt the choice of a  Lorentz general frame where the outgoing leptonic
current carries
$q^\mu = \left(q^+, q_\perp, q^- \right)= \left(\Delta P^+, q_\perp,
{q^2+q^2_\perp\over \Delta P^+}\right)$.  In the limit $\Delta \to 0$, the
matrix element for the $+$ vector current should coincide with the Drell-Yan
formula.

For the $n \to n$ diagonal term ($\Delta n = 0$),
the final-state hadron wavefunction has
arguments
$x_1-\Delta \over 1-\Delta$,
${\vec{k}}_{\perp 1} - {1-x_1\over 1-\Delta} {\vec{q}}_\perp$ for
the struck quark
and $x_i\over 1-\Delta$,
${\vec{k}}_{\perp i} + {x_i\over 1-\Delta} {\vec{q}}_\perp$
for the $n-1$ spectators.
We thus have a formula for the diagonal (parton-number-conserving) matrix
element of the form:
\begin{eqnarray}
{\VEV{A \vert J^\mu \vert B}}_{\Delta n = 0} &=&
\sum_{n, ~ \lambda}
\prod_{i=1}^{n} \int^1_{\Delta} dx_1
\int^1_0 dx_{i(i\ne 1)} \int {d^2{\vec{k}}_{\perp i} \over 2 (2\pi)^3 }
~ \delta\left(1-\sum_{j=1}^n x_j\right) ~ \delta^{(2)}
\left(\sum_{j=1}^n {\vec{k}}_{\perp j}\right)  \nonumber\\[1ex]
&&\times
\psi^\dagger_{A (n)}(x^\prime_i, {\vec{k}}^\prime_{\perp i},\lambda_i) ~ j ^\mu
~ \psi_{B (n)}(x_i, {\vec{k}}_{\perp i},\lambda_i),
\label{t1}
\end{eqnarray}
where
\begin{equation}
\left\{ \begin{array}{lll}
x^\prime_1 = {x_1-\Delta \over 1-\Delta}\, ,\
&{\vec{k}}^\prime_{\perp 1} ={\vec{k}}_{\perp 1}
- {1-x_1\over 1-\Delta} {\vec{q}}_\perp
&\mbox{for the struck quark}\\[1ex]
x^\prime_i = {x_i\over 1-\Delta}\, ,\
&{\vec{k}}^\prime_{\perp i} ={\vec{k}}_{\perp i}
+ {x_i\over 1-\Delta} {\vec{q}}_\perp
&\mbox{for the $ (n-1)$ spectators.}
\end{array}\right.
\label{t2}
\end{equation}
A sum over all possible helicities $\lambda_i$ is understood.
If quark masses are neglected the vector and axial currents conserve helicity.
 We also can check that $\sum_i^n x^\prime_i = 1$,
$\sum_i^n {\vec{k}}^\prime_{\perp i} = {\vec{0}}_\perp$.

For the $n+1 \to n-1$ off-diagonal term ($\Delta n = -2$),
let us consider the case where
partons $1$ and
$n+1$ of the initial wavefunction annihilate into the leptonic current leaving
$n-1$ spectators.
Then $x_{n+1} = \Delta - x_{1}$,
${\vec{k}}_{\perp n+1} = {\vec{q}}_\perp-{\vec{k}}_{\perp 1}$.
The remaining $n-1$ partons have total momentum
$((1-\Delta)P^+, -{\vec{q}}_{\perp})$.
The final wavefunction then has arguments
$x^\prime_i = {x_i \over (1- \Delta)}$ and
${\vec{k}}^\prime_{\perp i}=
{\vec{k}}_{\perp i} + {x_i\over 1-\Delta} {\vec{q}}_\perp$.
We thus obtain the formula for the off-diagonal matrix element:
\begin{eqnarray}
{\VEV{A\vert J^\mu \vert B}}_{\Delta n = -2} &=&
\sum_{n ~ \lambda}
\int^{\Delta}_0 dx_1 \int^1_0 dx_{n+1}
\int {d^2{\vec{k}}_{\perp 1} \over 2 (2\pi)^3 }
\int {d^2{\vec{k}}_{\perp n+1} \over 2 (2\pi)^3 }
\prod_{i=2}^{n}
\int^1_0 dx_{i} \int {d^2{\vec{k}}_{\perp i} \over 2 (2\pi)^3 }
\nonumber\\[2ex]
&&\times \delta\left(1-\sum_{j=1}^{n+1} x_j\right) ~
\delta^{(2)}\left(\sum_{j=1}^{n+1} {\vec{k}}_{\perp j}\right)
\nonumber\\[2ex]
&&\times\psi^\dagger_{A (n-1)}(x^\prime_i,{\vec{k}}^\prime_{\perp i},\lambda_i)
~ j^\mu
~ \psi_{B (n+1)}(\{x_1, x_i, x_{n+1} = \Delta - x_{1}\},
\nonumber\\[2ex]
&&\qquad\ \ \{ {\vec{k}}_{\perp 1},
{\vec{k}}_{\perp i},
{\vec{k}}_{\perp n+1} = {\vec{q}}_\perp-{\vec{k}}_{\perp 1}\},
\{\lambda_1,\lambda_{i},\lambda_{n+1} = - \lambda_{1}\}).
\label{t3}
\end{eqnarray}
Here $i=2,3,\cdots ,n$ with
\begin{equation}
x^\prime_i = {x_i\over 1-\Delta}\, ,\qquad
{\vec{k}}^\prime_{\perp i} ={\vec{k}}_{\perp i}
+ {x_i\over 1-\Delta} {\vec{q}}_\perp
\label{t3a}
\end{equation}
label the $n-1$ spectator
partons which appear in the final-state hadron wavefunction.
We can again check that the arguments of the final-state wavefunction
satisfy
$\sum_{i=2}^n x^\prime_i = 1$,
$\sum_{i=2}^n {\vec{k}}^\prime_{\perp i} = {\vec{0}}_\perp$.

The free current matrix elements $j^\mu$  in the light-cone
representation are easily constructed.  For example, the vector current of
quarks
takes the form
$$j^\mu =
{\bar u(x^\prime,k^\prime_\perp,\lambda^\prime)
\gamma^\mu
 u(x,k_\perp,\lambda) \over {\sqrt{k^+}} {\sqrt{k^{+ \prime}}}}$$
and  $$j^+ = 2 \delta_{\lambda , \lambda^\prime }\ .$$
The other light-cone spinor matrix elements  of
$j^\mu$ can be obtained from the
tables in ref. \cite{LB}.
In the case of spin zero partons
$$j^+ = {x + x^\prime \over \sqrt{ x  x^\prime}}$$
and
$$j^- =  {k^- + k^{\prime- }\over\sqrt { x x^\prime} P^+}.$$
However, instead of evaluating each $k^-$  in the $j^-$ current from the
on-shell
condition
$k^- k^+ = m^2$, one must instead  evaluate the $k^-$ of the struck
partons from
energy conservation
$k^- = p^-_{\rm initial} - p^-_{\rm spectator}$.  This effect is seen explicitly
when one integrates the covariant current over the denominator poles in the
$k^-$ variable.  It can also be understood as due to the implicit inclusion of
local instantaneous exchange contributions obtained in  light-cone
quantization \cite{CRY, BRS}.  The mass
$m_{\rm spectator}^2$ which is needed for the evaluation of
$j^-$ current in the diagonal case is the mass of the entire spectator system.
Thus
$m^2_{\perp {\rm spectator}} = m^2_{\rm spectator} +{\vec{k}}^2_{\perp {\rm
spectator}}$, where ${\vec{k}}_{\perp {\rm spectator}}= \sum_j {\vec{k}}_{\perp
j}$ and $m^2_{\perp {\rm spectator}}/x_{\rm spectator} = \sum_j m^2_j/x_j$,
summed
over the $j$ spectators.
This is an important simplification for  phenomenology, since we can change
variables to $m_{\rm spectator}^2$ and $d^2{\vec{k}}_{\perp {\rm
spectator}}$ and
replace all of the spectators by a spectral integral over the cluster mass
$m_{\rm spectator}^2$.  A specific example is presented in the next section.

\section{Example---\boldmath$\phi^3\ $\unboldmath Perturbation Theory}

As an explicit example and check on the above formalism, we shall consider the
electromagnetic vector current matrix element of a neutral composite system
composed of two charged scalars  where the light-cone wavefunctions are known
explicitly from  perturbation theory.  To construct the model, we consider a 3+1
dimensional system represented by the Lagrangian:
\begin{eqnarray}
{\cal{L}}&=&(\partial_\mu \phi_a+ie_aA_\mu\phi_a)^\dagger
(\partial^\mu \phi_a+ie_aA^\mu\phi_a ) -m_a^2\phi_a^\dagger\phi_a
\label{fm1}\\[1ex]
&& + (\partial_\mu \phi_b-ie_bA_\mu\phi_b)^\dagger
(\partial^\mu \phi_b-ie_bA^\mu\phi_b ) -m_b^2\phi_b^\dagger\phi_b
\nonumber\\[1ex]
&& +\half \partial_\mu\Phi \partial^\mu\Phi -\half M^2\Phi\Phi
+g \Phi (\phi_a^\dagger\phi_b+\phi_b^\dagger\phi_a ).
\nonumber
\end{eqnarray}
The composite system wavefunction can be normalized to unity by a choice of the
effective coupling $g$.
\vspace{.5cm}
\begin{figure}[htb]
\begin{center}
\leavevmode
\epsfbox{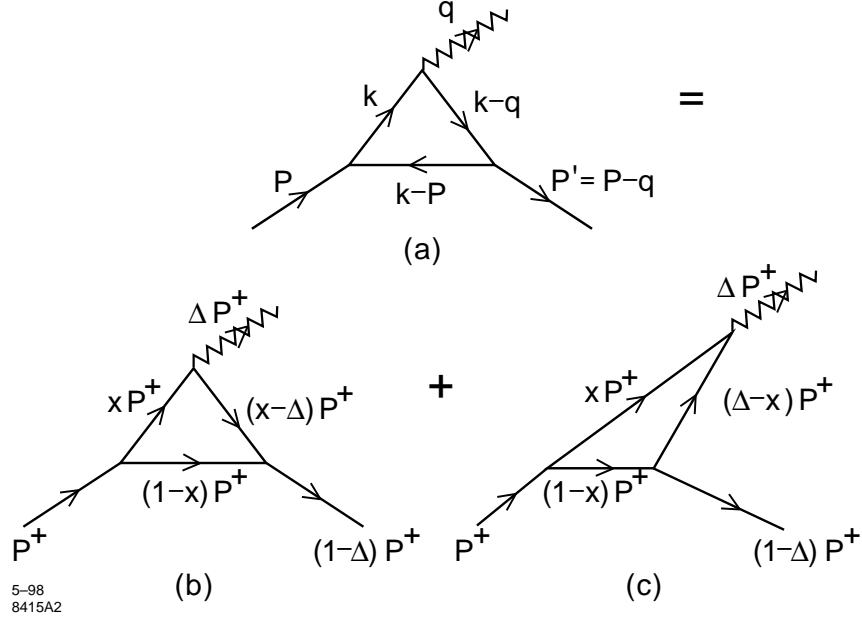}
\end{center}
\caption[*]{Scalar perturbation theory model for semileptonic decay. (a) Covariant
representation. (b), (c) Light-cone time-ordered contributions to the decay
amplitude.  These contributions can be identified as the  convolution of
light-cone Fock wavefunctions with $2 \to 2$ and
$3
\to 1$ parton number, respectively.
}
\label{fig2}
\end{figure}

We can derive the light-cone amplitudes from the covariant amplitude by
integrating over the $k^-$ variable \cite{MS}.
The amplitude of the process drawn in Fig. 2 is given as follows
from the Feynman rules:
\begin{eqnarray}
&&{\cal M}^\mu_a
\label{fa1}\\
&=&ie_ag^2\int {d^4k\over (2\pi)^4}\
{(2k-q)^\mu\over
(k^2-m_a^2+i\epsilon )\ ((k-q)^2-m_a^2+i\epsilon )((k-P)^2-m_b^2+i\epsilon )}
\nonumber\\[1ex]
&=&ie_ag^2\int {d^2{\vec{k}}_{\perp}\over 2(2\pi)^4}
\int P^+dx\ {1\over P^{+3}x(x-\Delta )(1-x)}
\nonumber\\[1ex]
&&\times\int dk^- {(2k-q)^\mu\over
\left(k^--{(m_a^2+{\vec{k}}_{\perp}^2)-i\epsilon\over xP^+}\right)
\left((k^--q^-)-{(m_a^2+({\vec{k}}_{\perp}-{\vec{q}}_{\perp})^2
-i\epsilon\over (x-\Delta )P^+}\right)
\left((k^--P^-)+{(m_b^2+{\vec{k}}_{\perp})^2)-i\epsilon\over (1-x)P^+}\right)},
\nonumber
\end{eqnarray}
where we used $k^+=xP^+$.
When we perform the integration over $k^-$, the integral does not vanish only
for $0\le x\le 1$.

\eject
For $\Delta \le x\le 1$, which corresponds to Fig. 2(b),
\begin{eqnarray}
&&({\cal M}^+_a,{\cal M}^-_a,{\vec{\cal M}}_{\perp a})_{(2 \to 2)}
\label{fa2} \\
&=&P^+e_ag^2\int_{\Delta}^1dx\int {d^2{\vec{k}}_{\perp}\over 2(2\pi)^3}
\nonumber\\
&&\times
\left((2x-\Delta )\ ,\ {1\over P^{+2}}
\left(2\left(M^2-{(m_b^2+{\vec{k}}_{\perp}^2)\over 1-x}\right)
-{(q^2+{\vec{q}}_{\perp}^2)\over\Delta}\right)
\ ,\ (2{\vec{k}}_{\perp}-{\vec{q}}_{\perp})\right)
\nonumber\\
&&\times
{1\over x(x-\Delta )(1-x)}
\nonumber\\
&&\times
{1\over \left(M^2-{m_a^2+{\vec{k}}_{\perp}^2\over x}
-{m_b^2+{\vec{k}}_{\perp}^2\over 1-x}\right)
\left({M^2+{\vec{q}}_{\perp}^2\over 1-\Delta}
-{m_a^2+({\vec{k}}_{\perp}-{\vec{q}}_{\perp})^2\over x-\Delta}
-{m_b^2+{\vec{k}}_{\perp}^2\over 1-x}\right)}
\nonumber\\
&=&P^+e_ag^2\int_{\Delta}^1dx\int {d^2{\vec{k}}_{\perp}\over 2(2\pi)^3}
\nonumber\\
&&\times
\left((2x-\Delta )\ ,\ {1\over P^{+2}}
\left(2\left(M^2-{(m_b^2+{\vec{k}}_{\perp}^2)\over 1-x}\right)
-{(q^2+{\vec{q}}_{\perp}^2)\over\Delta}\right)
\ ,\ (2{\vec{k}}_{\perp}-{\vec{q}}_{\perp})\right)
\nonumber\\
&&\times
{(1-\Delta )\over x(x-\Delta )(1-x)}
\nonumber\\
&&\times
{1\over \left(M^2-{m_a^2+{\vec{k}}_{\perp}^2\over x}
-{m_b^2+(-{\vec{k}}_{\perp})^2\over 1-x}\right)
\left(M^2
-{m_a^2+({\vec{k}}_{\perp}+(1-{x-\Delta\over 1-\Delta})(-{\vec{q}}_{\perp}))^2
\over {x-\Delta\over 1-\Delta}}
-{m_b^2+(-{\vec{k}}_{\perp}-{1-x\over 1-\Delta}(-{\vec{q}}_{\perp}))^2
\over {1-x\over 1-\Delta}}\right)}
\nonumber\\
&=&P^+e_ag^2\int_{\Delta}^1dx\int {d^2{\vec{k}}_{\perp}\over 2(2\pi)^3}
\nonumber\\
&&\times
{1\over {\sqrt{x(x-\Delta )}}}\left( (2x-\Delta )\ ,\ {1\over P^{+2}}
\left(2\left(M^2-{(m_b^2+{\vec{k}}_{\perp}^2)\over 1-x}\right)
-{(q^2+{\vec{q}}_{\perp}^2)\over\Delta}\right)
\ ,\ (2{\vec{k}}_{\perp}-{\vec{q}}_{\perp})\right)
\nonumber\\
&&\times
{1\over {\sqrt{x_ax_b}}}\
{1\over \left(M^2-{m_a^2+{\vec{k}}_{a\perp}^2\over x_a}\
-{m_b^2+{\vec{k}}_{b\perp}^2\over x_b}\right)}
\
{1\over {\sqrt{x'_ax'_b}}}\
{1\over \left(M^2-{m_a^2+{\vec{k'}}_{a\perp}^2\over x'_a}
-{m_b^2+{\vec{k'}}_{b\perp}^2\over x'_b}\right)}
\nonumber\\
&=&\int_{\Delta}^1dx\int {d^2{\vec{k}}_{\perp}\over 2(2\pi)^3}
\nonumber\\
&&
\qquad\times
\psi_{(2)}(x'_a,x'_b,{\vec{k'}}_{a\perp},{\vec{k'}}_{b\perp};M,m_a,m_b)
j_{(2 \to 2)a}^\mu
\psi_{(2)}(x_a,x_b,{\vec{k}}_{a\perp},{\vec{k}}_{b\perp};M,m_a,m_b),
\nonumber
\end{eqnarray}
where
\begin{eqnarray}
&&j_{(2 \to 2)a}=
e_a{P^+\over {\sqrt{x(x-\Delta )}}}
\label{fa2a}\\
&&\qquad\qquad\times
\left((2x-\Delta )\ ,\ {1\over P^{+2}}
\left( 2\left(M^2-{(m_b^2+{\vec{k}}_{\perp}^2)\over 1-x}\right)
-{(q^2+{\vec{q}}_{\perp}^2)\over\Delta}\right)
\ ,\ (2{\vec{k}}_{\perp}-{\vec{q}}_{\perp})\right),
\nonumber\\
&&\psi_{(2)}(x_a,x_b,{\vec{k}}_{a\perp},{\vec{k}}_{b\perp};M,m_a,m_b)=
g{1\over {\sqrt{x_ax_b}}}\
{1\over \left(M^2-{m_a^2+{\vec{k}}_{a\perp}^2\over x_a}
-{m_b^2+{\vec{k}}_{b\perp}^2\over x_b}\right)},
\nonumber\\
&&x_a=x,\quad x_b=1-x,\quad {\vec{k}}_{a\perp}={\vec{k}}_{\perp},\quad
{\vec{k}}_{b\perp}=-{\vec{k}}_{\perp},
\nonumber\\[1ex]
&&x'_a={x_a-\Delta\over 1-\Delta},\quad x'_b={x_b\over 1-\Delta},\quad
{\vec{k'}}_{a\perp}={\vec{k}}_{a\perp}+(1-x'_a){\vec{P'}}_{\perp},\quad
{\vec{k'}}_{b\perp}={\vec{k}}_{b\perp}-x'_b{\vec{P'}}_{\perp},
\nonumber\\[1ex]
&&{\vec{P'}}_{\perp}=-{\vec{q}}_{\perp}.
\nonumber
\end{eqnarray}

For $0\le x\le\Delta$, which corresponds to Fig. 2(c),
\begin{eqnarray}
&&({\cal M}^+_a,{\cal M}^-_a,{\vec{\cal M}}_{\perp a})_{(3 \to 1)}
\label{fa3}\\
&=&P^+e_ag^2\int_0^{\Delta}dx\int {d^2{\vec{k}}_{\perp}\over 2(2\pi)^3}
\nonumber\\
&&\times
\left((2x-\Delta )
\ ,\ {1\over P^{+2}}\left(2{(m_a^2+{\vec{k}}_{\perp}^2)\over x}
-{(q^2+{\vec{q}}_{\perp}^2)\over\Delta}\right)
\ ,\ (2{\vec{k}}_{\perp}-{\vec{q}}_{\perp})\right)
\nonumber\\
&&\times
{1\over
x(\Delta -x)(1-x)}
\nonumber\\
&&\times
{1\over \left(M^2-{m_a^2+{\vec{k}}_{\perp}^2\over x}
-{m_b^2+(-{\vec{k}}_{\perp})^2\over 1-x}\right)
\left(M^2-{m_a^2+{\vec{k}}_{\perp}^2\over x}
-{m_a^2+({\vec{q}}_{\perp}-{\vec{k}}_{\perp})^2\over \Delta -x}
-{M^2+{\vec{q}}_{\perp}^2\over 1-\Delta}\right)}
\nonumber\\
&=&
\int_0^{\Delta}dx\int {d^2{\vec{k}}_{\perp}\over 2(2\pi)^3}
\int_0^1dy\int {d^2{\vec{k}}_{y\perp}\over 2(2\pi)^3}
\nonumber\\[1ex]
&&\times
\psi_{(1)}\left({y\over 1-\Delta },{\vec{k}}_{y\perp}-{\vec{P'}}_{\perp}\right)
j_{(3\to 1)a}^\mu
\psi_{(3)}(x,y, \Delta -x,{\vec{k}}_{\perp},
{\vec{k}}_{y\perp},{\vec{q}}_{\perp}-{\vec{k}}_{\perp};M,m_a,m_b),
\nonumber
\end{eqnarray}
where ${\vec{P'}}_{\perp}=-{\vec{q}}_{\perp}$ and
\begin{eqnarray}
&&j_{(3\to 1)a}^\mu =e_a{P^+\over {\sqrt{x(\Delta -x)}}}
\nonumber\\
&&\qquad\times
\left((2x-\Delta )
\ ,\ {1\over P^{+2}}\left(2{(m_a^2+{\vec{k}}_{\perp}^2)\over x}
-{(q^2+{\vec{q}}_{\perp}^2)\over\Delta}\right)
\ ,\ (2{\vec{k}}_{\perp}-{\vec{q}}_{\perp})\right) ,
\label{fa5}\\
&&\psi_{(3)}(x,y, \Delta -x,{\vec{k}}_{\perp},
{\vec{k}}_{y\perp},{\vec{q}}_{\perp}-{\vec{k}}_{\perp};M,m_a,m_b)
\nonumber\\
&&\quad =
g^2{1\over {\sqrt{x(1-x)^2(\Delta -x)y}}}
\nonumber\\
&&\qquad\times
{1\over \left(M^2-{m_a^2+{\vec{k}}_{\perp}^2\over x}
-{m_b^2+(-{\vec{k}}_{\perp})^2\over 1-x}\right)
\left(M^2-{m_a^2+{\vec{k}}_{\perp}^2\over x}
-{m_a^2+({\vec{q}}_{\perp}-{\vec{k}}_{\perp})^2\over \Delta -x}
-{M^2+{\vec{k}}_{y\perp}^2\over y}\right)} ,
\nonumber\\[1ex]
&&\psi_{(1)}(y,{\vec{k}}_{y\perp})=
{{2(2\pi)^3\over \sqrt{y}}}\delta (y-1)
{\delta}^2({\vec{k}}_{y\perp}).
\nonumber
\end{eqnarray}

Then, by combining (\ref{fa2}) and (\ref{fa3})
the amplitude is given as:
\begin{eqnarray}
&&{\cal M}^\mu_a
\label{fa4}\\
&=&
\int_{\Delta}^1dx\int {d^2{\vec{k}}_{\perp}\over 2(2\pi)^3}\
\psi_{(2)}(x'_a,x'_b,{\vec{k'}}_{a\perp},{\vec{k'}}_{b\perp};M,m_a,m_b)
j_{(2 \to 2)a}^{\mu}
\psi_{(2)}\nonumber \\
&&\times (x_a,x_b,{\vec{k}}_{a\perp},{\vec{k}}_{b\perp};M,m_a,m_b)
\nonumber\\[1ex]
&&+
\int_0^{\Delta}dx\int {d^2{\vec{k}}_{\perp}\over 2(2\pi)^3}
\int_0^1dy\int {d^2{\vec{k}}_{y\perp}\over 2(2\pi)^3}
\nonumber\\[1ex]
&&\quad\times
\psi_{(1)}\left({y\over 1-\Delta },{\vec{k}}_{y\perp}-{\vec{P'}}_{\perp}\right)
j_{(3 \to 1)a}^{\mu}
\psi_{(3)}(x,y,\Delta -x,{\vec{k}}_{\perp},
{\vec{k}}_{y\perp},{\vec{q}}_{\perp}-{\vec{k}}_{\perp};M,m_a,m_b) .
\nonumber
\end{eqnarray}

By adding the above amplitude ${\cal M}^\mu_a$ and that given by exchanging
$a$ and $b$ ($e_a+e_b=0$), we obtain the total amplitude:
\begin{eqnarray}
&&{\cal M}^\mu =(2P-q)^\mu F(q^2)
\label{fa4a}\\[1ex]
&=&
\int_{\Delta}^1dx\int {d^2{\vec{k}}_{\perp}\over 2(2\pi)^3}\
\psi_{(2)}(x'_a,x'_b,{\vec{k'}}_{a\perp},{\vec{k'}}_{b\perp};M,m_a,m_b)
j_{(2 \to 2)a}^\mu
\psi_{(2)}\nonumber \\
&& \times (x_a,x_b,{\vec{k}}_{a\perp},{\vec{k}}_{b\perp};M,m_a,m_b)
\nonumber\\[1ex]
&&+
\int_0^{\Delta}dx\int {d^2{\vec{k}}_{\perp}\over 2(2\pi)^3}
\int_0^1dy\int {d^2{\vec{k}}_{y\perp}\over 2(2\pi)^3}
\nonumber\\[1ex]
&&\quad\times
\psi_{(1)}\left({y\over 1-\Delta },{\vec{k}}_{y\perp}-{\vec{P'}}_{\perp}\right)
j_{(3-1)a}^\mu
\psi_{(3)}(x,y,\Delta -x,{\vec{k}}_{\perp},
{\vec{k}}_{y\perp},{\vec{q}}_{\perp}-{\vec{k}}_{\perp};M,m_a,m_b)
\nonumber\\[1ex]
&&\qquad +\ \Bigl( a\leftrightarrow b\Bigr) ,
\nonumber
\end{eqnarray}
where ${\cal M}^\mu =(2P-q)^\mu F(q^2)$ follows from $q_\mu{\cal M}^\mu =0$.

For $q^2\rightarrow 0$, ${\vec{q}}_{\perp}\rightarrow {\vec{0}}_{\perp}$ and
$\Delta\rightarrow 0$,
$+$ component of (\ref{fa4a}) gives
\begin{eqnarray}
F(0)
&=&e_a\int {d^2{\vec{k}}_{\perp}\over (2\pi)^3}\int_0^1dx\
|\psi_{(2)}(x,1-x,{\vec{k}}_{\perp},-{\vec{k}}_{\perp};M,m_a,m_b)|^2
\label{fc1}\\
&&+e_b\int {d^2{\vec{k}}_{\perp}\over (2\pi)^3}\int_0^1dx\
|\psi_{(2)}(x,1-x,{\vec{k}}_{\perp},-{\vec{k}}_{\perp};M,m_b,m_a)|^2
\ =\ 0,
\nonumber
\end{eqnarray}
where
$\psi_{(2)}$  is given in (\ref{fa2a}).
Each term can be normalized to unit charge, thus providing
wavefunction renormalization in the model.  Alternatively we can evaluate the
$-$ component of (\ref{fa4a}) to obtain
\eject
\begin{eqnarray}
&&F(0)
\label{fc2}\\
&=&
e_a\int {d^2{\vec{k}}_{\perp}\over (2\pi)^3}
\, {1\over M^2}\, \Bigg( \int_0^1dx\
{1\over x}\, \left(M^2-{m_b^2+{\vec{k}}_{\perp}^2\over 1-x}
\right)
|\psi_{(2)}(x,1-x,{\vec{k}}_{\perp},-{\vec{k}}_{\perp};M,m_a,m_b)|^2\nonumber\\
&&+{1\over m_a^2+{\vec{k}}_{\perp}^2}g^2\Bigg)
\nonumber\\
&+&e_b\int {d^2{\vec{k}}_{\perp}\over (2\pi)^3}
\, {1\over M^2}\, \Bigg( \int_0^1dx\
{1\over x}\, \left(M^2-{m_a^2+{\vec{k}}_{\perp}^2\over 1-x}
\right)
|\psi_{(2)}(x,1-x,{\vec{k}}_{\perp},-{\vec{k}}_{\perp};M,m_b,m_a)|^2\nonumber\\
&& +{1\over m_b^2+{\vec{k}}_{\perp}^2}g^2 \Bigg) ,
\nonumber
\end{eqnarray}
where the ${e_a\over m_a^2+{\vec{k}}_{\perp}^2}g^2$
and ${e_b\over m_b^2+{\vec{k}}_{\perp}^2}g^2$ terms come from
the singular contributions of the
$\int_0^{\Delta}dx\, \psi_{(1)}\, j_{(3-1)}^-\, \psi_{(3)}$
terms in (\ref{fa4a}) when we take the limit $\Delta\rightarrow 0$.
The $\perp$ components of (\ref{fa4a}) do not give more information since
$(2{\vec{P}}-{\vec{q}})_{\perp}\rightarrow {\vec{0}}_{\perp}$
in the left hand side
and the integrand of the right hand side is odd about ${\vec{k}}_{\perp}$.

The above  analysis provides an explicit realization of the general formulas
(\ref{t1}) and (\ref{t3}).  In this simple model two transition matrix elements
appear: $ 2 \to 2 $ and $3 \to 1$.

The equality of the formulas for (\ref{fc1}) and (\ref{fc2}) is a general
condition which follows from gauge invariance and consistency of the
light-cone formalism.  We have verified the equality for the perturbative model
by direct evaluation of the integrals.

In the case of general composite systems, the equality of the form factors
at zero momentum transfer obtained from the $J^+$ and $J^-$ currents provides a
type of  virial theorem for the matrix elements (\ref{t1}) and (\ref{t3}).
In general the two determinations of the total charge $F(q^2=0)$  must be
consistent:
\begin{eqnarray}
F(0)&=&{1\over 2P^+}\,\, \lim_{q^2,{\vec{q}}_{\perp},\Delta\rightarrow 0}
{\VEV{A \vert J^+ \vert B}}_{\Delta n = 0},
\label{t123}\\
F(0)&=&{1\over 2P^-}\,\, \lim_{q^2,{\vec{q}}_{\perp},\Delta\rightarrow 0}
\left(
{\VEV{A \vert J^- \vert B}}_{\Delta n = 0}+
{\VEV{A \vert J^- \vert B}}_{\Delta n = -2}\right) .
\label{t123a}
\end{eqnarray}
Here $P^+ P^- = M^2_B.$  Note that the second term of (\ref{t123a}) includes the
zero mode
$\delta(x)$ contributions from the $n+1\to n-1$ off-diagonal matrix element.

\section{Conclusions}

A most important feature of the light-cone formalism is that all matrix
elements of local
operators can be written explicitly in terms of simple convolutions of
light-cone Fock
wavefunctions $\{\psi_n(x_i,k_{\perp i},\lambda_i)\}$.  In the case of
exclusive semileptonic $B$-decays, such as $B\rightarrow A \ell \bar{\nu}$,
the decay
matrix elements require the computation of  the diagonal matrix element
$n \rightarrow n$ where parton number is conserved and the off-diagonal
$n+1\rightarrow
n-1$ convolution where the current operator annihilates a $q{\bar{q'}}$ pair in
the initial $B$
wavefunction.  This term is a consequence of the fact that the time-like decay
$q^2 = (p_\ell + p_{\bar{\nu}} )^2 > 0$
requires a positive light-cone momentum fraction
$q^+ > 0$.  Conversely for space-like currents, one can choose $q^+=0$, as in
the Drell-Yan-West representation of the space-like electromagnetic form
factors.
However, as we have seen from the explicit analysis of the form factor in a
perturbation model, the off-diagonal convolution can yield a nonzero $q^+/q^+$
limiting form as $q^+
\rightarrow 0$.  This extra term appears specifically in the case of ``bad"
currents such
as $J^-$ in which the coupling to $q\bar q$ fluctuations in the light-cone
wavefunctions
are favored.  In effect, the $q^+ \rightarrow 0$ limit generates
$\delta(x)$ contributions as
residues of the $n+1\rightarrow n-1$ contributions.  The necessity for
this zero mode $\delta(x)$ terms were first noted in the pioneering work of
Chang, Root and Yan \cite{CRY},
and Burkardt analyzed it in his studies of higher-twist parton distributions
\cite{BUR}.
Here we see that the presence of such terms are
a general feature of local operator matrix elements when one selects the
simplified
$q^+=0$ frame.

We have also seen that the proper treatment of the $J^-$ current implies new
consistency conditions which must be obeyed by the light-cone wavefunctions.
For example, current conservation for the form factors of spin zero hadrons
requires
\begin{equation}
(2p-q)^\mu F(q^2) = \VEV{p-q\,|\,J^\mu(0)\, |\, p}
\end{equation}
and thus
\begin{equation}
\VEV{p-q\,|\,J^+\,|\,p} =
\frac{(2p-q)^+}{(2p-q)^-}\
\VEV{p-q\,|\,J^-\,|\,p} \ .
\end{equation}
We have explicitly verified this new type of virial theorem in a
simple scalar composite model in section 3.

The off-diagonal $n+1 \rightarrow n-1$ contributions provide a new
perspective on the
physics of $B$-decays.  A semileptonic decay involves not only matrix
element where a
quark changes flavor, but also a contribution where the leptonic pair is
created from the
annihilation of a $q {\bar{q'}}$ pair within the Fock states of the initial $B$
wavefunction.  The semileptonic decay thus can occur from the annihilation of a
nonvalence quark-antiquark pair in the initial hadron.  This feature will carry
over to exclusive hadronic
$B$-decays, such as
$B^0
\rightarrow
\pi^-D^+$.  In this case the pion can be produced from the coalescence of a
$d\bar u$ pair emerging from the initial higher particle number Fock
wavefunction of the $B$.  The $D$ meson is then formed from the remaining quarks
after the internal exchange of a $W$ boson.

We have emphasized the remarkable advantage of the light-cone formalism that all
matrix elements of local operators can be written down exactly in terms of
simple
convolutions of light-cone Fock wavefunctions.
The light-cone wavefunctions depend
only on the hadron itself; they are process-independent. The formalism is
relativistic and
frame-independent---the incident four-vectors can be chosen in any frame.
Note that the  matrix element of a current in the covariant Bethe-Salpeter
formalism requires the construction of the current from insertions into an
infinite number of irreducible kernels.  The Bethe-Salpeter formalism becomes
even more intractable for bound-states of more than two particles.

In principle, a precise evaluation of the hadronic
matrix elements needed for $B$-decays
and other exclusive electroweak decay amplitudes requires knowledge of all of
the light-cone Fock wavefunctions of the initial and final state hadrons.
In the case of some model gauge theories such as QCD \cite {Horn}  or collinear
QCD \cite {AD} in one-space and one-time dimensions, the complete evaluation of
the light-cone wavefunction is possible for each baryon or meson bound-state
using the discretized light-cone quantization method.  It would be
interesting to
use such solutions as a model for physical $B$-decays.

The evaluation of
the light-cone Fock wavefunctions in QCD(3+1) is not at present
computationally feasible because of the large number of degrees of freedom
within
the hadron wavefunctions. Nevertheless, the existence of an exact formalism
provides a basis for systematic approximations and a control over neglected
terms.  For example, one can analyze exclusive semileptonic
$B$-decays  which  involve a heavy internal momentum transfer using a
perturbative QCD formalism patterned after the analysis of form factors at
large momentum transfer \cite{LB}.    The hard-scattering analysis proceeds by
writing each hadronic wavefunction as a sum of soft and hard contributions
\begin{equation}
\psi_n = \psi^{{\rm soft}}_n (\M^2_n < \Lambda^2) + \psi^{{\rm hard}}_n
(\M^2_n >\Lambda^2) ,
\end{equation}
where
\begin{equation}
\M^2_n = \sum^n_{i=1} \left(\frac{k^2_\perp+m^2}{x}\right)^2_i
\end{equation}
is the invariant mass of the partons in the $n$-particle Fock state and
$\Lambda$ is the
separation scale.
The high internal momentum contributions to the wavefunction $\psi^{{\rm
hard}}_n $ can be calculated systematically from QCD perturbation theory
from the
interaction of the gluon exchange kernel.  The contributions from  high
momentum transfer exchange to the
$B$-decay amplitude can then be written as a convolution of a hard scattering
quark-gluon scattering amplitude $T_H$  with the distribution
amplitudes
$\phi(x_i,\Lambda)$, the valence wavefunctions obtained by integrating the
constituent momenta  up to
the separation scale
${\cal M}_n < \Lambda < Q$.  This is the basis for the perturbative hard
scattering
analyses of Refs. \cite {BHS, Sz, BALL, BABR}. In our exact analysis, one can
identify the hard PQCD contribution as well as the soft contribution from
the convolution of the light-cone wavefunctions. Furthermore, the hard
scattering contribution can be systematically improved.  For example, off-shell
effects can be retained in the evaluation of
$T_H$ by utilizing the exact light-cone energy denominators.
This effect will be analyzed in
a separate paper.
\\

\begin{center}
{\bf Acknowledgements}
\end{center}

\noindent
We wish to thank Chueng-Ryong Ji, Yong-Yeon Keum, Bum-Hoon Lee, and
Adam Szczepaniak for helpful discussions.
This work was supported
in part by the Basic Science Research Institute Program,
Ministry of Education, Project No. BSRI-97-2414,
and in part by Non-Directed-Research-Fund,
Korea Research Foundation 1997.\\

\vfill\eject

\end{document}